\documentclass{iaus}
\usepackage{graphicx}
\usepackage{color}
\usepackage{subfig}

\title[DNN for studies of galaxy morphology] 
{Deep learning for studies of galaxy morphology}

\author[D. Tuccillo, et al.]   
{D. Tuccillo$^{1,2}$
 \thanks{E-mail:diego.tuccillo@obspm.fr},
 M. Huertas-Company,$^{1}$
 E. Decenci\`ere,$^{2}$
\and S. Velasco-Forero,$^{2}$}

\affiliation{$^1$GEPI, Observatoire de Paris, CNRS, Universit\'e Paris Diderot,  \\61, Avenue de l'Observatoire F-75014, Paris, France \\
$^2$ MINES ParisTech, PSL Research University, CMM \\ Centre for mathematical morphology, Fontainebleau, France \\}

\pubyear{xxx}
\volume{xxx}  
\jname{IAU Symposium 325 on Astroinformatics}
\editors{A.C. Editor, B.D. Editor \& C.E. Editor, eds.}
\begin{document}

\maketitle

\begin{abstract}
Establishing accurate morphological measurements of galaxies in a reasonable amount of time for future big-data surveys such as EUCLID, the Large Synoptic Survey Telescope or the Wide Field Infrared Survey Telescope is a challenge. Because of its high level of abstraction with little human intervention, deep learning appears to be a promising approach. Deep learning is a rapidly growing discipline that models high-level patterns in data as complex multilayered networks. In this work we test the ability of deep convolutional networks to provide parametric properties of Hubble Space Telescope like galaxies (half-light radii, S\'ersic indices, total flux etc..). We simulate a set of galaxies including point spread function and realistic noise from the CANDELS survey and try to recover the main galaxy parameters using deep-learning. We compare the results with the ones obtained with the commonly used profile fitting based software GALFIT. This way showing that with our method we obtain results at least equally good as the ones obtained with GALFIT but, once trained, with a factor 5 hundred time faster. 
\end{abstract}

\firstsection 
\section{Introduction}

%

Deep Neural Networks (DNN) are a particular kind of artificial neural network architectures that allow a high level of abstraction thanks to the use of many layers and neurons. They are particularly suited for problems where the target function is very complex and the datasets are very large. Most of the theory on DNN was developed in the 1980s and 1990s, thanks to the foundation work of \cite{Fukushima_1980} and \cite{LeCun_1989}. However, initially, they did not found great success being applied to real problems.  Recently (\cite{Krizhevsky_2012}), also thank to the use of powerful graphics processing unit (GPU) and larger datasets, DNNs have finally attracted wide-spread attention  by outperforming alternative machine learning methods such as kernel-based algorithms. Especially in problems of pattern and speech recognition. Among others \cite{Schmidhuber_2015} gives a thorough review of the field.

In astronomy DNN have been proven to be effective to build automated separation of star/galaxies (\cite{Kim_2017}) and as classification systems for galaxy morphologies, a task where they have obtained an accuracy greater than $\sim 90\%$ to predict various aspects of galaxy morphology, using directly raw pixel data without feature extraction.  Such accuracy has been obtained both in the work from \cite{Dieleman_2015} using data from the Galaxy Zoo project and at high redshift in \cite{Huertas_2015} for CANDELS fields (\cite{Galametz_2013}) galaxies. 

Two-dimensional profile fitting is often used to measure the properties of galaxies .  Most of the known algorithms used in astronomy use a technique called parametric fitting. They use  least-squares algorithm of non-linear type to adjust the light profile of the galaxy with analytic functions commonly used in astronomy to model the galaxy morphology, like the S\'ersic, exponential models. The parameters are chosen searching for best fit minimizing the $\chi ^2$. If the fit is successful, the essential features of the object are estimated in few parameters like its luminosity, its radial size and light central concentration. The task of fitting functions to a galaxy image often needs careful attentions to judge whether a fit is physically meaningful, and to use the words from \cite{Peng_2002}: requires some element of scientific, technical, and not infrequently, artistic, sense. These algorithms can be very reliable and accurate for small samples of galaxies but they are not conceived to deal with extremely large amount of data. Nevertheless, an automated approach for this kind of tasks is becoming unavoidable as modern telescopes continue to collect more and more images every day. Future surveys like EUCLID, the Large Synoptic Survey Telescope (LSST) and the Wide Field Infrared Survey Telescope (WFIRST) , will vastly increase the number of galaxy images that can be morphologically analyzed. As a result, the classical methods to classify and analyze them cannot be expected to scale indefinitely with the growing amount of data.  

In this work we present the first tests of a new approach to the problem, consisting in developing a DNN machine to automatically estimate galaxy luminosities and their main parameters from 2-D light profiles images as given from optical surveys.  If successful,  the new method should be able to provide accurate morphological measurements significantly faster than existing techniques. 
   
\section{Galaxy surface brightness profile fitting}

We approached the problem of two-dimensional galaxy profile fitting, developing a Deep Neural  Network algorithm to estimate  the parameters of a single - S\'ersic profile. We built a sample of simulated Hubble Space Telescope (HST) CANDELS  galaxies (section \ref{simulatedSec}), to train and validate our DNN machine.  In section \ref{gamoclass} we describe our algorithm and in section \ref{compaGALFIT} we compare its performance with GALFIT applied to the same set of simulations. In section \ref{twoCompGalaxies} we show preliminary results on the use of our DNN to fit two component galaxies.

\subsection{Simulated data}
\label{simulatedSec}

We simulate a set of 31,000 galaxies. We use the F160W (H band) Point spread function (PSF) from the HST and real noise from the CANDELS survey (\cite{Koekemoer_2013}). Our stamps have a size of $128 \times 128$ pixels and a pixel scale of 0.06''.  The parameters follow a uniform random distribution (logarithmic for the radius) for the structural parameters, having:

\begin{itemize}
\item $ 0<Radius  \, (arcsec)<1.9$
\item $18 < Magnitude  \, (AB) < 23$
\item $0.2 < Ellipticity < 0.8$
\item $0.3 <$ S\'ersic  index $< 6.3$
\item $0<Position  \, Angle  \, (degree)<180$
\end{itemize}

%

\subsection{Architecture of the DNN} 
\label{gamoclass}

We used 30,000 of the simulated stamps described in the previous subsection to train and validate different architectures of DNN to infer all the structural parameters used to simulate the galaxies. The 4/5  of the sample i.e. 24,000 stamps of galaxies, were used to train the DNNs and 1/5, i.e. 6,000 to validate them. The best architecture of the DNN was chosen  as the one minimizing the validation loss, i.e., the average loss function on the validation sample. 1000 additional galaxies were used to test the trained machine and compare to GALFIT.

We developed our DNN as a convolutional neural network (CNN) implemented in the Keras framework on top of Theano  (\cite{Bastien_2012}), two frameworks commonly used to build Deep Learning models.

\begin{figure}
\centering
\includegraphics[width=0.95 \textwidth]{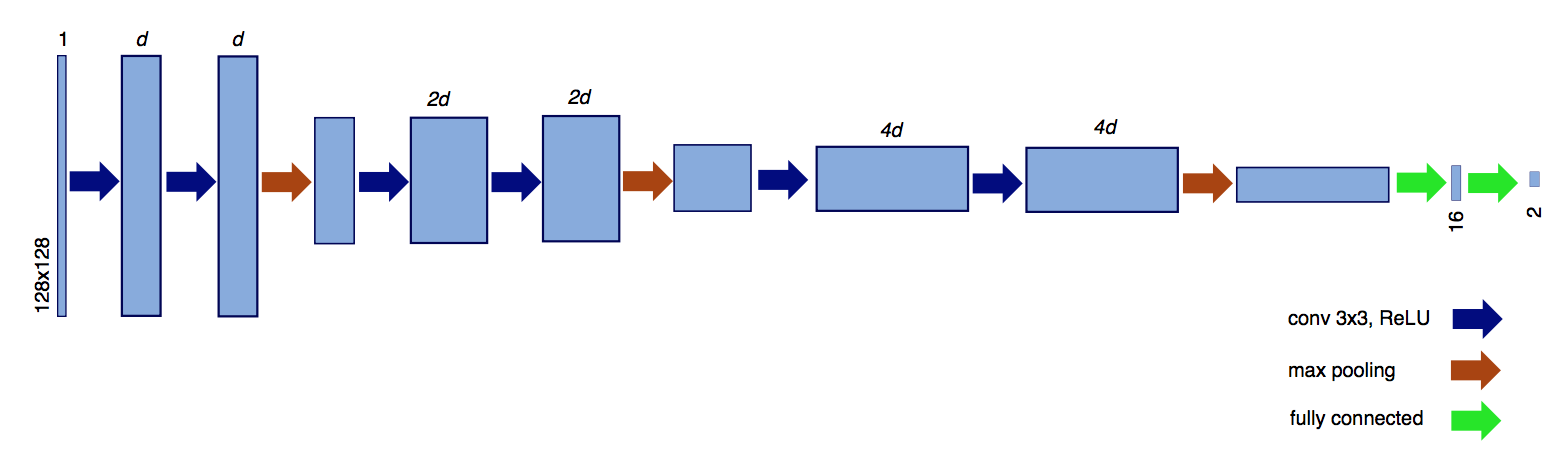}
\caption{Schematic representation of the CNN that we used in the work here presented.  It comprises 6 convolutional layers. Three max-pooling layers and 2 fully connected layers.}
\label{DNNArch_fig}
\end{figure}

Our best CNN is schematically shown in Fig. \ref{DNNArch_fig} and comprises six convolutional layers with filter of sizes $3\times3$ and $2 \times  2$. Each convolutional layer is followed by a rectified linear unit (ReLU) step. Three max-pooling layers with window size of $2 \times 2$ ,  two of them followed by a  "dropout layer". By dropping a unit out, we
mean temporarily removing it from the network. The final layer is composed of 2 fully connected layers. We fitted one parameter at time, so the number of neurons in the second fully connected layer is equal to one. No data argumentation  was used in the experiments. For weights updates was used a the Stochastic Gradient Descend (SGD) optimization method.

\subsection{Performance and comparison with GALFIT} 
\label{compaGALFIT}

\begin{figure}
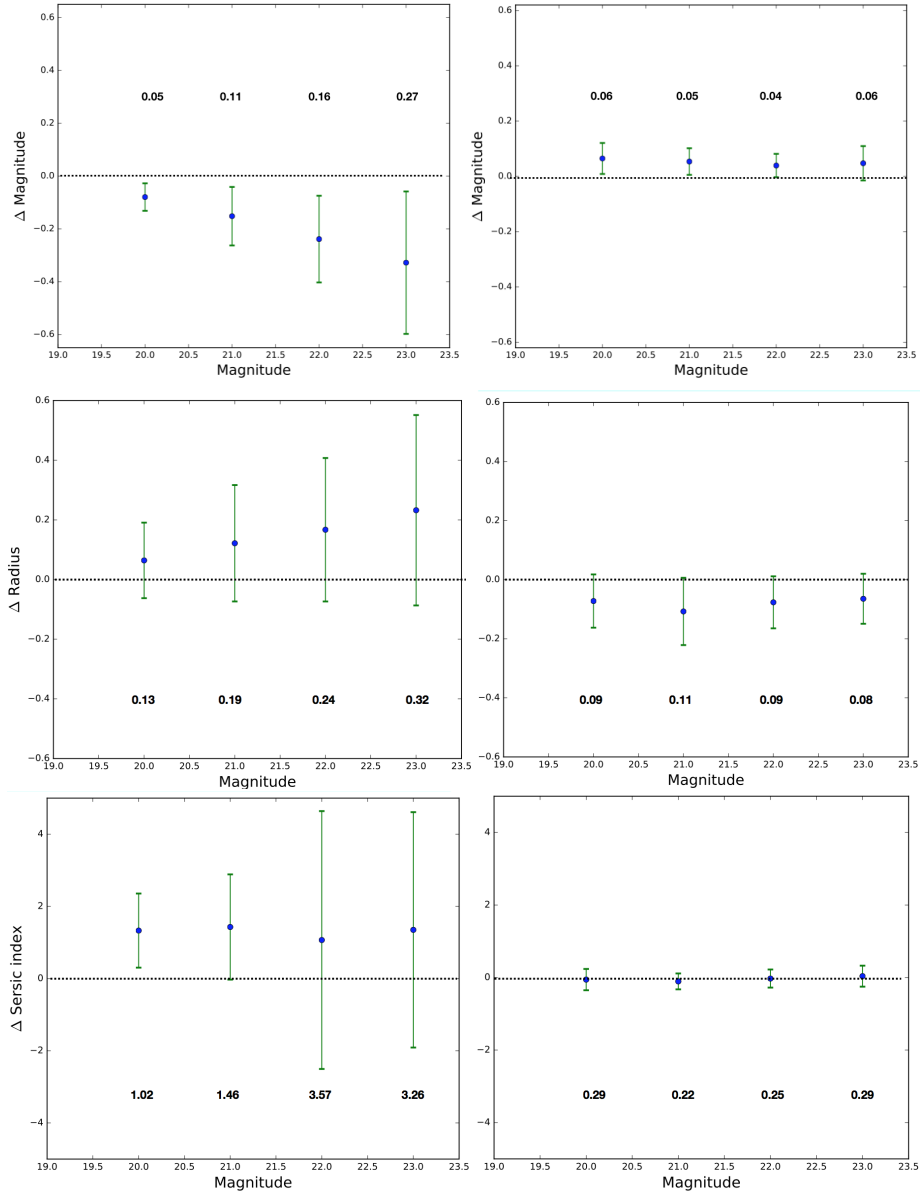

\centering
\includegraphics[width=0.9 \textwidth]{magScatter.pdf}\\
\includegraphics[width=0.9 \textwidth]{scatRad.pdf}\\
\includegraphics[width=0.9 \textwidth]{scaSer}\\
\caption{Comparison of of GALFIT and our CNN on predicting various parameters of 1000 HST/CANDELS simulated galaxies. The left column shows the GALFIT results and the right column the ones from the CNN. The comparison is made looking at the difference between the value of the parameters used to simulate the galaxy and the value obtained from the fitting. We compare the mean and the standard deviation of this quantities in bins of the total magnitude used to simulate the galaxies.  The first rows shows the mean and scatter in the recovered total magnitude, the second row show them for the radius of the galaxy and the last row for the S\'ersic index.}
\label{Fig2}
\end{figure}

Once that the weights of our best CNN model were fixed with the training, the resulting model was applied to 1000 stamps of simulated-galaxies that we previously excluded from the training and validation. We fitted the same sample with GALFIT, using standard setting used in \cite{VanDerWel_2012}. Thus we compared the predictions obtained with these two softwares. For all the parameters (radius of the galaxy, magnitude, ellipticity, S\'ersic index) we obtained comparable or better results than GALFIT in terms of both bias and dispertion. In Fig \ref{Fig2} we show in particular the comparison for the magnitude, radius and S\'ersic index. The comparison is made looking at the difference between the value of the parameters used to simulate the galaxy and the value obtained from the fitting. We compare the mean and the standard deviation of this quantities in bins of the total magnitude used to simulate the galaxies. A complete description of the CNN software and of of its performance in predicting all the galaxy parameters will be given in Tuccillo et al. (2016, in prep). In that paper we will also discuss the application of the CNN to fit 2-D images of real galaxies.

On the computational side we notice that GALFIT takes about 40 minutes to fit these 1000 galaxies, while our trained CNN takes only a few seconds.  The CNN have to be trained just one time and it takes about 20 minutes to be trained on 30,000 galaxies.

\subsection{Further tests: results on two component galaxies} 
\label{twoCompGalaxies}

\begin{figure}%
\centering
\subfloat{{\includegraphics[width=6cm]{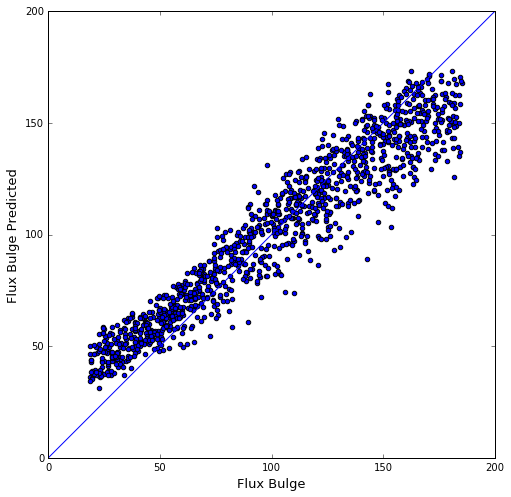} }}%
\qquad
\subfloat{{\includegraphics[width=6cm]{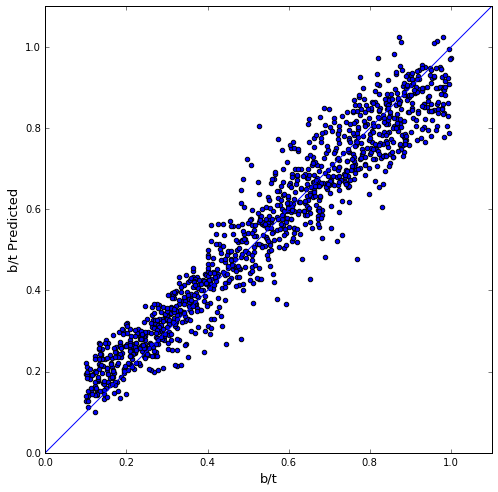} }}%
\caption{Results of our CNN in fitting the bulge flux and the b/t ratio of two component galaxies. On the x-axis we plot the parameter used to simulate the galaxy and on the y-axis the parameter fitted by our algorithm}%
\label{2componentRes}%
\end{figure}

We also tested the performance of our CNN on fitting the parameters of 2-components galaxies. Following the same procedure, we simulated the images 30,000 double S\'ersic HST/CANDELS galaxies, and we used them to train and validate our CNN. We did not change the architecture of the neural network. We estimated the structural parameters of both disk and bulge, so for each of this components: magnitude, radius, S\'ersic index and ellipticity. We also estimated global parameters such as the b/t ratio, the total magnitude and the total radius of the galaxies. The results on the validation sample are very promising although not yet as good as the ones described in the previous section for the one component galaxies. We are now working to change the CNN architecture in order to obtain better results and understand the cases in which the CNN parameter-estimation is very different from the value used to simulate the galaxy.  As example we show in Fig. \ref{2componentRes} the results obtained for fitting the bulge flux and the b/t ratio.

\section{Discussion - Overview}

We presented in this work the first results of a CNN developed to fit the galaxy parameters of one-component galaxy, and we show at least comparable results  than those obtained using  GALFIT on the same test-sample. We also show preliminary results of profile fitting of simulated two component galaxies.   We are currently working on the optimization of our CNN and on the problem of \textit{transfer learning}, i.e. the application of the CNN model optimized on simulated data on real data retrieved from astronomical surveys.  

\begin{figure}
\centering
\includegraphics[width=0.4 \textwidth]{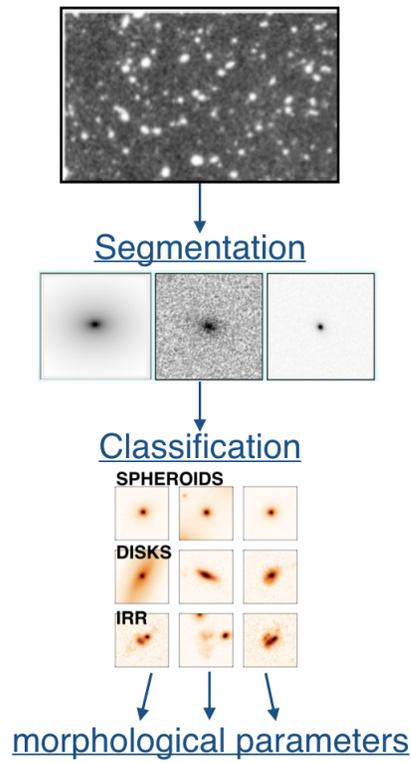}\\
\caption{The first column correspond to the target image of the validation set, i.e. the ideal output of the segmentation,  the second column correspond to the images of the validation set, and finally the third column correspond to the result of the segmentation performed by the DNN}
\label{Fig6}
\end{figure}

The results presented here are part of a more ambitious project consisting in developing a full deep neural network machine to automatically label and fit galaxies on CCD images produced from large area surveys. The full logical outflow of our project is presented schematically in Fig.\ref{Fig6}. Given an image from a CCD we aim to: 1) segment the image in order to separate galaxies from the rest of the objects; 2) classify the galaxy on the basis if their morphological type like already done successful by our investigation group in \cite{Huertas_2015}; 3) fit the parameters of the galaxies depending on their morphological type.

\begin{thebibliography}{}

\bibitem[Bastien et al., 2012]{Bastien_2012}
{Bastien F. et al.,} (2012), 
\textit{NIPS}, 2012

\bibitem[Dieleman, Willett, \& Dambre (2015)]{Dieleman_2015}
{Dieleman, S., Willett, K. W., \& Dambre, J.} 2015, 
\textit{MNRAS}, 450, 1441

\bibitem[Fukushima (1980)]{Fukushima_1980}
{Fukushima, K.} 1980,
\textit{Biological Cybernetics}, 36, 193-202 

\bibitem[Galametz et al. 2013]{Galametz_2013}
{Galametz, A., Grazian, A., Fontana, A., et al.} 2013, 
\textit{ApJS}, 206, 10

\bibitem[Huertas-Company  et al. (2015)]{Huertas_2015}
{Huertas-Company M., et al.} 2015, 
\textit{ApJS}, 221, 8

\bibitem[Kim \&Brunner, 2017]{Kim_2017}
{Kim, E. \& Brunner, R.} 2017, 
\textit{MNRAS}, 464, 4463-4475


\bibitem[Koekemoer  et al., 2013]{Koekemoer_2013}
{Koekemoer A. M. et al.} 2013, 
\textit{ApJS}, 209, 3

\bibitem[Krizhevsky, Sutskever, \& Hinton, 2012]{Krizhevsky_2012}
{Krizhevsky, A., Sutskever, I. \& Hinton, G.} 2012, 
\textit{NIPS}1097-1105

\bibitem[LeCun et al. (1989)]{LeCun_1989}
{LeCun,Y.,  Boser, B., Denker, J. S., Henderson, D., Howard, R. E., Hubbard, W., Jackel,  L. D.} 1989, 
\textit{Neural Computation}, 1 (4):541-551

\bibitem[Peng et al. (2002)]{Peng_2002}
{Peng, C.~Y. et al.} 2002, 
\textit{AJ}, 124, 266


\bibitem[Schmidhuber (2015)]{Schmidhuber_2015}
{Schmidhuber, J.} 2015,
\textit{Neural Netw}, 61, 85?117



\bibitem[van der Wel, et al. (2012)]{VanDerWel_2012}
{van der Wel A., et al.} (2012)
\textit{ApJS}, 203, 24









\end{thebibliography}
\end{document}